\RequirePackage{fix-cm}
\documentclass[twocolumn]{svjour3}      % twocolumn
\journalname{Granular Matter}
\smartqed  % flush right qed marks, e.g. at end of proof
\usepackage{graphicx}
\usepackage{bm}
\usepackage{mathptmx}      % use Times fonts if available on your TeX system
\usepackage{amsbsy}
\usepackage{color}
\usepackage{xcolor}
\usepackage{amsmath}
\begin{document}\sloppy

\title{Evolution of Internal Granular Structure at the Flow-Arrest Transition}


\author{Ishan Srivastava \and Jeremy B. Lechman \and Gary S. Grest \and Leonardo E. Silbert}


\institute{I. Srivastava \and J.B. Lechman \and G.S. Grest \at Sandia National Laboratories, Albuquerque, New Mexico 87185, USA, \\\email{isriva@sandia.gov}
               \and L.E. Silbert \at School of Math, Science, and Engineering, Central New Mexico Community College, Albuquerque, New Mexico 87106, USA \\\email{lsilbert@cnm.edu}}

\date{}

\maketitle

\begin{abstract} 
  The evolution of the internal granular structure in shear-arrested and
  shear-flowing states of granular materials is characterized using fabric
  tensors as descriptors of the internal contact and force networks. When a
  dilute system of frictional grains is subjected to a constant pressure and
  shear stress, the bulk stress ratio is well-predicted from the
  anisotropy of its contact and force networks during transient flow prior to steady shear flow or shear arrest. Although the onset of shear arrest is a stochastic process, the fabric tensors upon arrest are distributed around nearly equal contributions of force and contact network anisotropy to the bulk stress ratio. The distribution becomes seemingly narrower with increasing system size. The anisotropy of the contact network in shear-arrested states is reminiscent of the fabric anisotropy observed in shear-jammed packings.
  
\end{abstract}

\section{Introduction}
Granular materials respond to applied stress in several remarkable ways. When
subjected to a critical ratio of shear stress and pressure, they can
plastically deform with accompanying dilation to reach a \emph{critical state}
beyond which further quasistatic deformation is volume preserving and rate
independent~\cite{schofield1968critical}. When the applied stress is
sufficiently large, granular materials exhibit inertial frictional rheology,
which is described by a proportionality between the internal shear stress and
rate of shear deformation~\cite{Jop:2006gh,Salerno:2018en}. Therefore, a
well-defined friction-dependent critical ratio of shear stress and pressure
demarcates the plastic or fluid-like deformation of granular materials from a
static solid-like behavior~\cite{singh2013effect,Srivastava:2019bc}. In
addition to stress, the solids volume fraction of granular materials also
controls their mechanical behavior: below a friction-dependent critical volume
fraction (also known as the jamming transition), granular materials display
fluid-like behavior, whereas above this volume fraction, the grains are jammed
and collectively exhibit mechanical
rigidity~\cite{Liu:2010jx,o2003jamming,Srivastava:2019bc}. However, this
picture has been complicated by recent observations that granular materials
can jam below the so-called jamming volume fraction by subjecting them to external
shear
strain~\cite{Bi:2011bt,vinutha2016disentangling,Chen:2018cf,clark2018critical}.

The application of shear to a granular material can introduce significant
geometrical anisotropy within its microstructure. When sheared at its
boundaries, the constituent grains tend to align preferentially in the
direction of shear~\cite{Majmudar:2005bn}. Such directional alignment of the
grains also results in the partial alignment of the force network giving rise
to the formation of directional force chains---as depicted in
Figs.~\ref{fig1}(b) and (c)---which carry a majority of the shear
stress~\cite{Radjai:1998hn}. Consequently, several metrics have been proposed
to quantify structural anisotropy in granular materials, of which the most
commonly utilized is a second order symmetric fabric
tensor~\cite{Radjai:2012fc,Rothenburg:1989iq}. Early work by Rothenburg and
Bathurst~\cite{Bathurst:1990bw} demonstrated that the shear stress within
granular materials can be partitioned into contributions from various fabric
tensors, thereby providing a direct connection between bulk mechanics and
inherent granular structure. Several later studies have developed continuum
models for granular plasticity and rheology incorporating the granular fabric
tensor as a key internal variable~\cite{Radjai:2012fc,SUN:2011ks}. The effect
of structural anisotropy on the bulk mechanical response is even more
pronounced when granular materials are subjected to complex loading paths.

Prior studies have comprehensively explored the evolution of fabric anisotropy
during quasi-static deformation of granular materials leading to critical
state and failure~\cite{Guo:2013gs}. The structural anisotropies in granular
materials during steady inertial flows as a function of applied shear rate
have also been demonstrated~\cite{Azema:2014ej}. However, the transient
evolution of granular fabric during its transitions between solid-like and
fluid-like states remains largely unexplored. Such studies will especially aid
the development of advanced constitutive laws that can predict time-dependent
state of stress and deformation in granular materials as they transition
between fluid-like and solid-like behaviors.

Recent simulations have demonstrated that the stress-induced transient
flow-to-arrest transition in frictional granular materials is a highly
stochastic process with a very wide distribution of arrest times, along with
power-law divergences of arrest statistics near a critical shear stress, all of
which are sensitive to interparticle friction~\cite{Srivastava:2019bc}. In
this work we elucidate the transient evolution of the granular fabric during
such a flow-arrest transition. We provide evidence that stress partitioning
into various fabric components---which has been successfully applied to
quasistatic deformation~\cite{Guo:2013gs} and steady inertial
flows~\cite{Azema:2014ej}---remains valid even during the transient flow-arrest
transition. We highlight that despite the stochastic nature of this
transition, the structural anisotropy of the granular material upon arrest
remains largely invariant to these statistics, indicating the presence of a
unique shear-arrested state for a given loading protocol.

\section{Simulations}
Dilute granular systems---composed of $N$ slightly dispersed spheres (within $5$\% of average diameter $d$)---are prepared at an initial volume fraction $\phi\!=\!0.05$. The contact mechanics between spheres is modeled using a damped Hookean spring-dashpot viscoelastic model, along with tangential Coulomb friction characterized by an interparticle friction coefficient $\mu_s$~\cite{Silbert.2001}. The tangential spring stiffness $k_t$ is set equal to the normal spring stiffness $k_n\!=\!1$. The normal velocity damping coefficient $\gamma_n=0.5$ and the tangential velocity damping coefficient is $\gamma_t=0.5\gamma_n$. In the present simulations, time is normalized by the characteristic timescale $\sqrt{m/k_n}$ where $m$ is the mass of a particle of diameter $d$ and density $=1$, and the simulation time step is set to $0.02\sqrt{m/k_n}$. The energy and stress in the simulations are scaled by $k_n d^2$ and $k_n/d$ respectively.

A constant state of stress at the periodic system boundaries is maintained
using the modularly-invariant dynamical equations~\cite{shinoda2004rapid} of the Parrinello-Rahman (PR) method~\cite{Parrinello:1981it} within
LAMMPS~\cite{Plimpton:1995fc}, which allows the triclinic periodic box to deform in all possible ways, i.e., shear, dilation/compaction, and rotation, in response to external stress. The equations of motion associated with the dynamics of periodic box are described in the Appendix. The novelty of the present simulation method is that, unlike strain-controlled simulations, it enables simulating granular states in the vicinity of the flow-arrest transition along paths of constant pressure and shear stress, without any complications associated with boundary effects~\cite{silbert2002boundary,schuhmacher2017wall}.  

\begin{figure}[t!]
\includegraphics[width=\columnwidth]{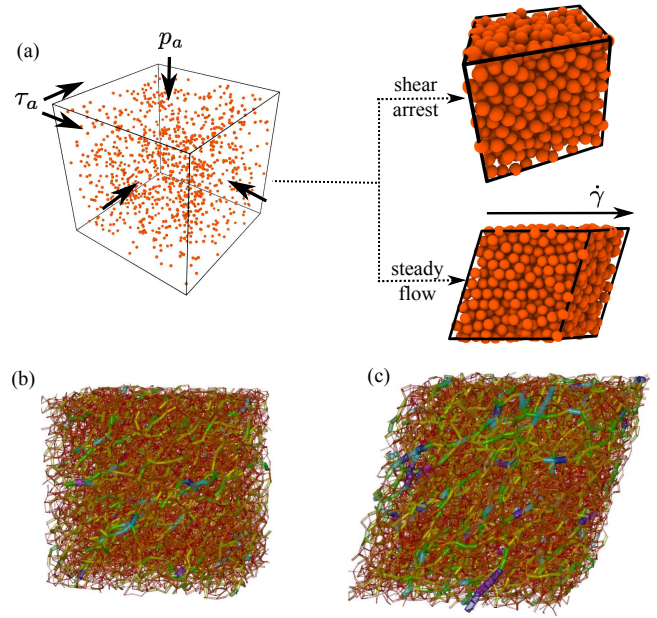}
\caption{(a) Schematic of the simulation method. The left image represents a starting configuration at low density $\phi\!=\!0.05$, subjected to external pressure $p_{a}$ and shear stress $\tau_{a}$. The images on the right are representative of the final, compacted state, which shear arrest or steady flow at a constant strain rate $\dot{\gamma}$, depending on the value of $\tau_{a}/p_{a}$. (b) and (c) respectively depict force networks within a granular system at early and late times during a representative simulation. The magnitude of contact forces between particles are color-coded (reds and yellows corresponds to smaller forces whereas blues are largest) and also represented by the diameter of the connecting cylinder.}
\label{fig1}
\end{figure}

We simulated granular systems with two values of interparticle friction $\mu_{s}\!=\!0.001$ and $1.0$ representing regimes of
low and high friction respectively. More than $500$ distinct starting states for $N\!=\!10^{4}$ and more than $100$ distinct starting states for
$N\!=\!3\times10^{3}$ were simulated to provide robust statistics for
shear-arrested states, following the observations in
Ref.~\cite{Srivastava:2019bc} that the time for the onset of arrest in a
flowing granular material is highly unpredictable. We also performed
simulations for frictionless particles and particles with intermediate  friction $\mu_{s}\!=\!0.065$, and observed similar evolution of the internal granular structure at the flow-arrest transition.

The initial dilute systems are subjected to a constant applied shear stress
$\tau_{a}$ and applied pressure $p_{a}$ (denoted by the subscript $a$), such that the total applied stress $\bm{\sigma}_{a}$ follows: (i) $(1/3)\!\sum\sigma_{a,ii}\!=\!p_{a}$,
(ii) $\sigma_{a,ij}\!=\!\tau_{a}$ for $i,j\!=\!1,\!2$ and $2,\!1$, and (iii) $\sigma_{a,ij}\!=\!0$ for all other indices $i\!\neq\!j$. While the pressure is fixed at $p_{a}\!=\!10^{-4}k_{n}/d$ to simulate hard particle regime at
high volume fraction, $\tau_{a}$ is varied to simulate two distinct
long-time scenarios: (i) steady shear flow at a constant strain rate
$\dot{\gamma}$, and (ii) shear arrest where $\dot{\gamma} \to 0$. Here the scalar strain rate $\dot{\gamma}$ is computed as $\sqrt{\frac{1}{2}\bm{\dot{\gamma}}_{D}:\bm{\dot{\gamma}}_{D}}$, where $\bm{\dot{\gamma}}_{D}$ is the deviatoric component of the total 3D strain rate tensor $\bm{\dot{\gamma}}$, which is calculated from the deformation of the triclinic periodic box. The simulation method is described schematically in Fig.~\ref{fig1}(a).

In the present analyses, the internal (Cauchy) stress state $\bm{\sigma}$ of the system is tracked during its evolution under external applied stress from the interparticle contact forces as: $\bm{\sigma} =
\sum_{c}\bm{f}_{c}\otimes \bm{r}_{c}$, where $\bm{f}_{c}$ and $\bm{r}_{c}$ are the force and branch vectors between two contacting particles, and the
summation is over all contacts~\cite{thompson2009}. The stresses are defined negative in the \emph{compressive} sense. During long times prior to and at the flow-arrest transition, the kinetic contribution to the total internal stress is insignificant compared to contact stress contribution, and it is ignored in the present analyses. Two relevant scalar invariants of Cauchy stress are measured, (i) pressure
$p\!=\!(1/3)\!\sum\sigma_{ii}$ and (ii) shear
$\tau\!=\!\sqrt{\frac{1}{2}\bm{\tau}:\bm{\tau}}$, where $\bm{\tau}$ is the deviatoric component of the total internal stress. Consequently, the bulk stress ratio $\mu$ of a granular material is defined as:
\begin{equation}
    \mu=\tau/p
\label{eq0}
\end{equation}

\section{Bulk Response}
\begin{figure}[t!]
\includegraphics[width=\columnwidth]{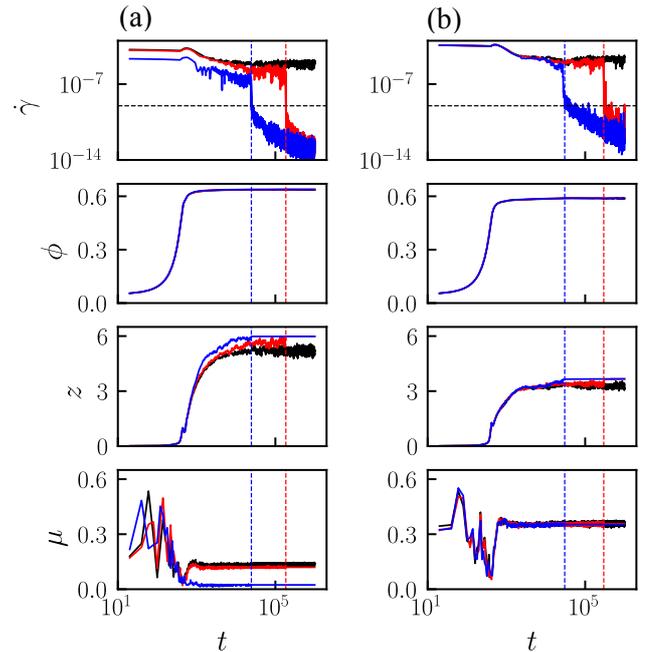}
\caption{From top to bottom: variation of strain rate $\dot{\gamma}$, volume fraction $\phi$, coordination number $z$ and bulk stress ratio $\mu$ with time $t$ for three values of applied stress ratio $\tau_{a}/p_{a}$ for interparticle frictions (a) $\mu_{s}\!=\!0.001$, and (b) $\mu_{s}\!=\!1.0$. Blue, red and black curves in (a) and (b) represent evolution for increasing $\tau_{a}/p_{a}$ respectively, where black represents steady flowing state, and blue and red represent shear arrest (see main text for values of $\tau_{a}/p_{a}$). Vertical dashed lines of the same color indicate the time for the onset of arrest $t_{c}$, while the horizontal dashed line in the upper panels represent the criterion for determining the onset of arrest.}
\label{fig2}
\end{figure}

Figure~\ref{fig2} demonstrates the evolution of key bulk variables, i.e., strain rate $\dot{\gamma}$, volume fraction $\phi$, average coordination number $z$ and bulk stress ratio $\mu$ for the same starting state of $N\!=10^{4}$ particles subjected to different values of the applied stress ratio $\tau_{a}/p_{a}$. Two distinct dynamical scenarios emerge: (i) steady shear flow at a constant mean $\dot{\gamma}$ for high applied shear $\tau_{a}/p_{a}\!=\!0.137$ for low interparticle friction (black lines in Fig.~\ref{fig2}(a)) and $\tau_{a}/p_{a}\!=\!0.357$ for high interparticle friction (black lines in Fig.~\ref{fig2}(b)); and (ii) shear arrest at a distinct time $t_{c}$ for low shear $\tau_{a}/p_{a}\!=\!0.019$ (blue) and $0.119$ (red) for low friction (Fig.~\ref{fig2}(a)), and $\tau_{a}/p_{a}\!=\!0.348$ (blue) and $0.351$ (red) for high friction (Fig.~\ref{fig2}(b)). The shear arrest is followed by $\dot{\gamma} \to 0$ for $t>t_{c}$. The onset time for shear arrest $t_{c}$ is defined as the time at which $\dot{\gamma}$ drops below a minimal value of $10^{-9}$, as indicated by dashed vertical lines in Fig.~\ref{fig2}. Although the onset of arrest---on average---occurs at longer times for higher applied stress ratio $\tau_{a}/p_{a}$ below a critical value, there is large variability in $t_{c}$ for a given stress ratio depending on the initial granular configuration, as described in~\cite{Srivastava:2019bc}. 

At early times the granular system compacts to a friction-dependent volume fraction, which remains invariant with time for both shear arrest and shear flow. For low interparticle friction, $\phi\!\sim\!0.64$, which is similar to the random close pack density of frictionless spheres, while $\phi\!\sim\!0.59$ for high friction, which is similar to the critical volume fraction in the \emph{critical state} plasticity theory~\cite{schofield1968critical}. Although the bulk strain rate exhibits a bifurcation of several orders of magnitude depending upon the applied shear stress, volume fraction is insensitive to this transition.

Unlike $\phi$, the coordination number $z$ exhibits differences between long-time steady shear flow and shear arrest. The early time compaction of the granular system is accompanied by a commensurate rise in $z$, after which it fluctuates around a mean value during a period of transient flow at a constant volume, prior to steady shear flow or arrest. The mean value of $z$ during transient flow is dependent on both shear stress and friction, with a moderately larger $z$ for lower shear stress and lower friction, consistent with previous observations on steady inertial flows~\cite{Azema:2014ej}. While $z$ continues to fluctuate around a mean value for systems entering steady shear flow, the onset of shear arrest is characterized by a sudden change in $z$ to a shear-independent value regardless of the onset of arrest, as shown in Fig.~\ref{fig2}. For low friction particles $z\!=\!5.98\!\pm\!0.01$, whereas $z\!=\!3.63\!\pm\!0.04$ for high friction particles. Hence, upon arrest these values are similar to the coordination number obtained for isotropic jamming of frictional and frictionless particles~\cite{o2003jamming,Silbert.2010,Otsuki:2011et}.  These estimates for $z$ were averaged from multiple shear arrest simulations, and the errors represent the standard deviation. These results demonstrate the uniqueness of coordination number for a given interparticle friction for all shear-arrested states attained through the same loading protocol regardless of arrest statistics and applied shear stress.

Finally, the bulk stress ratio $\mu$ (bottom panels in Fig.~\ref{fig2}) show a dependence on not only the applied stress but also the particle friction coefficient. However, despite these numerical differences, the focus of the next section is to elucidate the role that fabric plays in contributing to the overall $\mu$ and how these different systems exhibit similar behavior once the system has achieved its steady or arrested state.

\section{Fabric Evolution}
Volume fraction and coordination number describe the mechanical behavior of
granular materials in the vicinity of isotropic fluid to solid jamming
transition, as observed by the scaling of these quantities with system
pressure~\cite{Goodrich:2016ch}. However, the anisotropy induced within
contact and force networks in granular materials upon the application of shear stress requires additional characterization beyond these isotropic
measures. Fabric tensors provide a directional representation of the contact and force networks in granular materials, and consequently, the inherent
structural anisotropy within these networks~\cite{KenIchi:1984gt,ODA:1982dt}. The geometry of a
contact network can be described by the distribution $P(\vec{n})$ of
orientation of contact normal $\vec{n}$ in terms of a fabric tensor:
\begin{equation}
R_{ij} = \frac{N_{c}}{V}\int_{V} P(\vec{n}) n_{i} n_{j} \textrm{d} \vec{n},
\label{eq1}
\end{equation}
where $n_{i}$ is the $i$-th component of $\vec{n}$, $N_{c}$ is the number of contacts, and $V$ is the volume of the system. Numerically, $R_{ij}$ is computed from the simulation data through the following relation:
\begin{equation}
    R_{ij}=\frac{1}{N_{c}}\sum_{N_{c}}n_{i}n_{j},
    \label{eq2}
\end{equation}
where the summation is over all the contacts. It is often convenient to express the contact orientation distribution in terms of a second-order Fourier expansion of $P(\vec{n})$~\cite{Rothenburg:1989iq,Bathurst:1990bw}: 
\begin{equation}
P(\vec{n}) = \frac{1}{4\pi}\left[1+a^{c}_{ij}n_{i}n_{j}\right],
\label{eq3}
\end{equation}
where $a^{c}_{ij}$ is a symmetric and deviatoric \emph{contact anisotropy tensor} whose magnitude represents geometrical anisotropy within the contact network, and which is related to the fabric tensor by $a^{c}_{ij}=\frac{15}{2}R^{'}_{ij}$, where $R^{'}_{ij}$ is the deviatoric part of $R_{ij}$.

Similarly, the force-carrying network in a granular material presents a source of mechanical anisotropy, which are quantified using tensors representing the force network. Particularly, a \emph{normal force anisotropy tensor} $a^{n}_{ij}$ and a \emph{tangential force anisotropy tensor} $a^{t}_{ij}$ are defined such that orientational distribution of normal forces $f^{n}(\vec{n})$ and tangential forces $f^{t}(\vec{n)}$ are represented in their second-order Fourier expansion~\cite{Bathurst:1990bw}:
\begin{eqnarray}
f^{n}(\vec{n}) &=& f_{0} \left[1+a^{n}_{ij}n_{i}n_{j}\right], \nonumber \\
f_{i}^{t}(\vec{n}) &=&  f_{0} \left[a^{t}_{ik}n_{k}-(a^{t}_{kl}n_{k}n_{l})n_{i}\right],
\label{eq4}
\end{eqnarray}
where $f_{0}$ is the mean normal contact force within the granular material.

\begin{figure}[t!]
\includegraphics[width=\columnwidth]{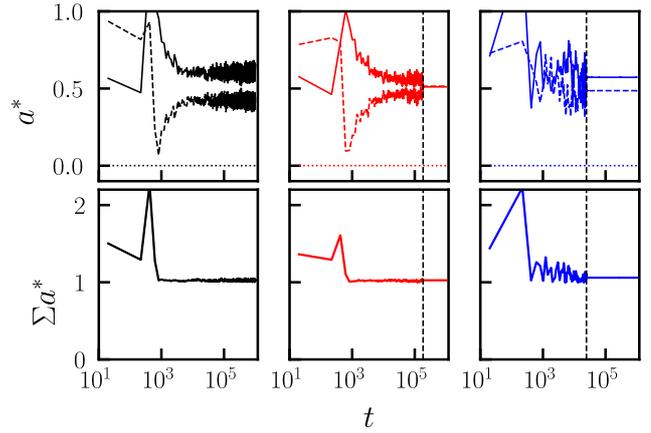}
\caption{(Top row) From left to right: evolution of scaled fabric anisotropies: contact $a_{c}^{*}$ (solid), normal force $a_{n}^{*}$ (dashed), and tangential force $a_{t}^{*}$ (dotted) with time $t$ for the three cases displayed in Fig.~\ref{fig2}(a) for interparticle friction $\mu_{s}\!=\!0.001$. Blue, red and black curves correspond to simulations of the same color displayed in Fig.~\ref{fig2}(a). (Bottom row) Evolution of the sum of scaled fabric anisotropies $\sum a^{*}$ with $t$ for the three cases in the top row. Black vertical dashed lines indicates the time for the onset of arrest $t_{c}$ for the two cases the arrest (second and third columns).}
\label{fig3}
\end{figure}

Prior studies have shown that the internal stress in granular materials can be partitioned into orientational distributions of contact normal and force vectors as follows~\cite{kanatani1981theory,Rothenburg:1989iq,Bathurst:1990bw}: 
\begin{equation}
\sigma_{ij}=\frac{N_{c}l_{0}}{V}\int_{\Omega} f_{i}(\vec{n}) n_{j} P(\vec{n}) \textrm{d}\vec{n},
\label{eq5}
\end{equation}
where the integration is over the entire angular domain $\Omega$, $l_{0}$ is the mean center-to-center distance of contacting particles, and $f_{i}(\vec{n})$ is the orientational distribution of the total contact force vector, which can be decomposed into normal and tangential contributions, as shown in Eq.~\ref{eq4}. Combining Eqs.~\ref{eq3},~\ref{eq4} and~\ref{eq5} along with the definition $\mu$ in Eq.~\ref{eq0} reveals the following first-order stress-force-fabric relation~\cite{Guo:2013gs}:
\begin{equation}
\mu \simeq \frac{2}{5}\left(a_{c}+a_{n}+\frac{3}{2}a_{t}\right),
\label{eq6}
\end{equation}
where $a_{c}$, $a_{n}$ and $a_{t}$ are the second invariants of the deviatoric fabric tensors $a^{c}_{ij}$, $a^{n}_{ij}$ and $a^{t}_{ij}$ respectively. The equality in Eq.~\ref{eq6} holds true only if the three fabric tensors are co-axial with the stress tensor. Regardless of the applied stress or interparticle friction, we observe co-axiality of all fabric tensors with the stress tensor at all times except during early transients when the contact network is still evolving.

\begin{figure}[t!]
\includegraphics[width=\columnwidth]{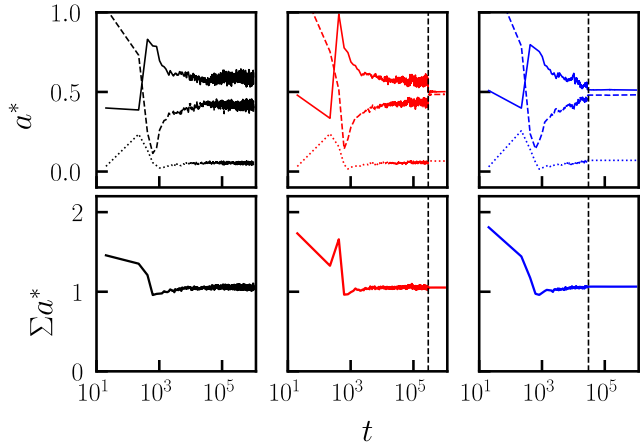}
\caption{(Top row) From left to right: evolution of scaled fabric anisotropies: contact $a_{c}^{*}$ (solid), normal force $a_{n}^{*}$ (dashed), and tangential force $a_{t}^{*}$ (dotted) with time $t$ for the three cases displayed in Fig.~\ref{fig2}(b) for interparticle friction $\mu_{s}\!=\!1.0$. Blue, red and black curves correspond to simulations of the same color displayed in Fig.~\ref{fig2}(b). (Bottom row) Evolution of the sum of scaled fabric anisotropies $\sum a^{*}$ with $t$ for the three cases in the top row. Black vertical dashed lines indicates the time for the onset of arrest $t_{c}$ for the two cases the arrest (second and third columns).}
\label{fig4}
\end{figure}

Figure~\ref{fig3} demonstrates the evolution of three scaled fabric anisotropies $a_{c}^{*}\!=\!\left(2/5\right)a_{c}/\mu$, $a_{n}^{*}\!=\!\left(2/5\right)a_{n}/\mu$, and $a_{t}^{*}\!=\!\left(3/5\right)a_{t}/\mu$ for the same bulk response depicted in Fig.~\ref{fig2}(a) (same color) for low friction at different values of applied $\tau_{a}$. After the early transients during which the contact and force networks develop as a result of compaction, the stress-force-fabric relation in Eq.~\ref{eq6}, i.e., $\sum a^{*}\!=\!a_{c}^{*}\!+\!a_{n}^{*}\!+\!a_{t}^{*}\!=\!1$, is satisfied at all times even during the transient flowing state before the system either arrests (red and blue curves) or continues flowing indefinitely in a steady flow (black curves), as shown in lower panels of Fig.~\ref{fig3}. This equality also demonstrates the co-axiality of the stress tensor with all the fabric tensors. However, the relative contribution of fabric anisotropies to the bulk stress ratio varies significantly with time. Early in the transient flowing state, contact anisotropy $a_{c}$ dominates the total system anisotropy, whereas the force network remains fairly isotropic. This indicates that although a large number of contacts begin to align in the direction of shear, rapid breaking and forming of new contacts does not sustain the directionality of the force network. As time progresses, the relative contribution of contact (force) network decreases (increases) towards seemingly steady values, in which $a_{c}$ and $a_{n}$ provide nearly equal contribution to the bulk stress ratio. Figs.~\ref{fig1}(b) and (c) respectively depict the force network for a representative simulation of low friction particles during early and late times, and the isotropic (anisotropic) nature of the force network at early (late) times is apparent. The trends in the evolution of fabric anisotropies are similar for particles with high friction, as shown by the variation of $a_{c}^{*}$, $a_{n}^{*}$ and $a_{t}^{*}$ in Fig.~\ref{fig4}, corresponding to same bulk response depicted in Fig.~\ref{fig2}(b). Similar to particles with low friction, the stress-force-fabric relation in Eq.~\ref{eq6} is satisfied at all times except early transients, with a larger anisotropy of the tangential force network that is still significantly small compared to contact and normal force anisotropies. The near-equality of $a_{c}$ and $a_{n}$ for quasistatic flows near the flow-arrest transition is consistent with previous observations of fabric evolution in steady inertial flows~\cite{Azema:2014ej}.

The fabric anisotropies of steady flowing systems fluctuate around a mean value at long times. For systems that arrest, the three fabric anisotropies instantaneously become time-invariant, indicating an arrested microstructure. Regardless of interparticle friction or applied shear, $a_{c}^{*}\! \approx \!a_{n}^{*}\! \approx \!0.5$ upon arrest, indicating that shear-arrested states are uniquely defined by friction-dependent volume fraction, coordination number and fabric anisotropies, with nearly equal contribution of contact and force network anisotropies. It is expected that shear-arrested fabric anisotropies depend significantly upon the loading protocol, which is same for all systems in the present simulations.

\begin{figure}[t!]
\includegraphics[width=\columnwidth]{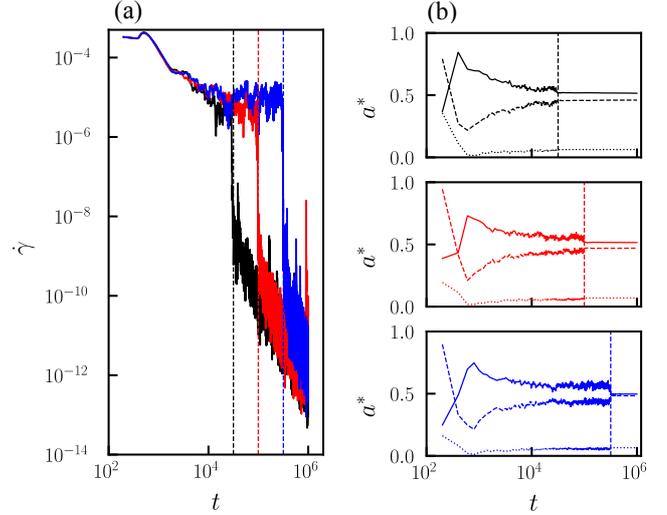}
\caption{(a) Evolution of strain rate $\dot{\gamma}$ for three different starting states at the same applied stress ratio $\tau_{a}/p_{a}=0.351$ for interparticle friction $\mu_{s}=1.0$. (b) Evolution of scaled fabric anisotropies: contact $a_{c}^{*}$ (solid), normal force $a_{n}^{*}$ (dashed), and tangential force $a_{t}^{*}$ (dotted) for each of the three cases displayed in (a). The vertical dashed lines indicate the onset of arrest $t_{c}$ for each case.}
\label{fig5}
\end{figure}

\section{Fabric States at Shear Arrest}
Recent simulations have demonstrated that the flow to arrest transition in frictional granular materials is a highly stochastic process. The onset of arrest below critical shear at time $t_{c}$ exhibits a heavy-tailed distribution, whose moments diverge near the critical shear threshold~\cite{Srivastava:2019bc}. As an example, Fig.~\ref{fig5}(a) demonstrates shear arrest for three different starting states of granular systems subjected to the same sub-critical shear stress $\tau_{a}$, and with the same interparticle friction. Although all the three systems eventually arrest, their onset of arrest times $t_{c}$ vary by more than an order of magnitude. Here we examine whether this stochasticity associated with shear arrest is reflected in the response of the contact and force fabric of the granular material. 

The evolution of three scaled fabric anisotropies for three cases of shear arrest in Fig.~\ref{fig5}(a) is shown in Fig.~\ref{fig5}(b). The response of the fabric after early transients and before shear arrest is qualitatively equivalent in all the three cases: at early times, $a_{c}^{*}$ dominates the bulk stress ratio, and as time progresses, $a_{c}^{*}$ and $a_{n}^{*}$ evolve to nearly equivalent values. However, more crucially, the state of the fabric upon arrest is nearly the same for all the three cases, with $a_{c}^{*}\! \approx \!a_{n}^{*}\! \approx \!0.5$, thereby indicating that the fabric is independent of the time to arrest. 

\begin{figure}[t!]
\includegraphics[width=\columnwidth]{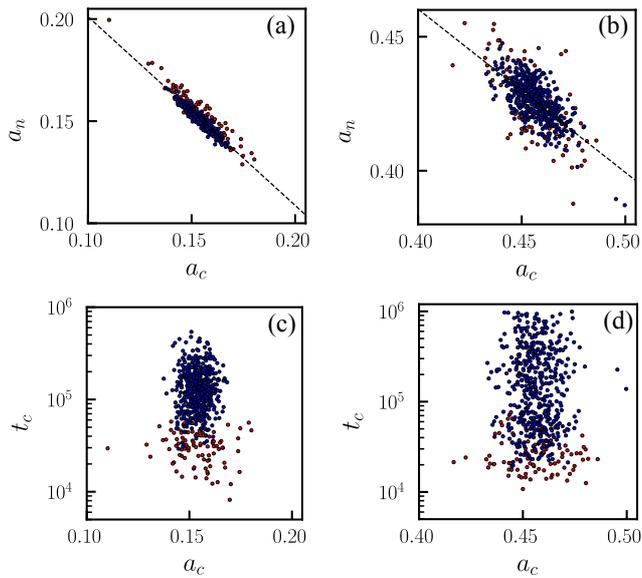}
\caption{Correlation between normal force fabric anisotropy $a_{n}$ and contact fabric anisotropy $a_{c}$ for (a) interparticle friction $\mu_s=0.001$ and applied stress ratio $\tau_{a}/p_{a}=0.119$, and (b) interparticle friction $\mu_s=1.0$ and applied stress ratio $\tau_{a}/p_{a}=0.351$. Dashed lines represent the best linear fit with slopes $-0.92$ and $-0.61$ for (a) and (b) respectively.  Scatter of the time for the onset of arrest $t_{c}$ and contact fabric anisotropy $a_{c}$ in (c) and (d) correspond to the data in (a) and (b) respectively. Points in (a) - (d) represent data from system sizes $N\!=\!3\!\times\!10^{3}$ (red) and $N\!=\!10^{4}$ (blue).}
\label{fig6}
\end{figure}

We examined the statistics of the state of the fabric upon arrest in greater detail by simulating multiple starting states towards shear arrest for the same value of applied shear. Figs.~\ref{fig6}(a) and (b) respectively highlight the scatter of $a_{c}$ and $a_{n}$ for all the cases of shear arrest for low friction and high friction respectively. The fabric anisotropies are scattered around (non-scaled) mean values of $a_{c}\!=\!0.154\!\pm\!0.005$, $a_{n}\!=\!0.150\!\pm\!0.005$ and $a_{t}\!=\!6.9\!\times\!10^{-5}\!\pm\!4\!\times\!10^{-6}$ for low friction, and $a_{c}\!=\!0.456\!\pm\!0.009$, $a_{n}\!=\!0.426\!\pm\!0.007$ and $a_{t}\!=\!0.060\!\pm\!0.003$ for high friction. Expectedly, higher interparticle friction results in greater microstructural anisotropy which also leads to a larger bulk stress ratio through the stress-force-fabric relation. For smaller system sizes, the scatter of the fabric anisotropies is larger, as shown in Fig.~\ref{fig6}, although they are scattered around a similar mean value. This signifies the importance of finite system size effects, and it also possibly indicates that for an infinitely large system, the distribution will vanish around unique values of fabric anisotropies. A systematic study of system size effects will better elucidate the uniqueness of arrested fabric. 

The narrow distribution of the fabric anisotropies indicates the uniqueness of the fabric at shear arrest. Unlike $a_{c}$ and $a_{n}$ which are slightly negatively correlated for finite system sizes (see dashed linear correlation fits in Figs.~\ref{fig6}(a) and (b)) due to the constraints of Eq.~\ref{eq6}, we find no correlation between arrest times $t_{c}$ and fabric anisotropies. Figs.~\ref{fig6}(c) and (d) highlight the scatter of $t_{c}$ and $a_{c}$, and no correlation between the two is observed, thereby implying that no matter how long the granular system takes to arrest from its flowing state, it will arrest to a uniquely defined internal granular structure. 

\section{Conclusions}
The results presented here have two important implications. Shear jamming of granular materials was observed in experiments~\cite{Bi:2011bt} and simulations on frictional~\cite{Otsuki:2011et} and frictionless~\cite{Chen:2018cf,clark2018critical,vinutha2016disentangling} particles through various strain-controlled protocols. These studies observed \emph{fragile} states in addition to shear-jammed states, and a theory was developed to identify the transition between them~\cite{Sarkar:2013fj}. The shear-arrested states simulated in this study through a completely stress (and pressure)-controlled protocol bear resemblance to those observed in previous studies. They are highly anisotropic in their fabric, and importantly, their internal state of stress is uniquely defined by the stress invariants $\mu$ and $p$, unlike strain-controlled methods where the internal stress is not uniquely defined~\cite{Chen:2018cf}. Further analysis is required to highlight similarities and differences between such shear-arrested and shear-jammed states of granular matter. Additionally, it remains to be seen whether these shear-arrested states exhibit signs of fragility~\cite{Bi:2011bt}, i.e., any directional instability to shear. Recent stress-controlled simulations on frictional suspensions at a constant volume provide insight into the differences between shear-jammed and fragile states of granular materials~\cite{Seto2019}. 

Recent advances in the continuum modeling of granular materials include anisotropic effects in constitutive models~\cite{Radjai:2017bl}. Particularly, advances to the classical critical state theory have included the role of fabric~\cite{zhao2013unique,Li:2012kb,Li:2015ka,Dafalias:2016gh}---in addition to volume fraction and stress ratio---to uniquely define a critical state of granular plasticity at which granular materials deform indefinitely at a constant volume and stress. The uniqueness of critical state granular fabric has also been backed by recent simulations~\cite{zhao2013unique,Wang:2017ch}. The simulations presented here highlight the \emph{opposite} phenomenon of the arrest of a flowing granular material. Currently, no constitutive models exist that can predict transient granular flow prior to shear arrest, and the stochastic nature of shear arrest itself. However, the preceding results, which elucidate the unique state of granular fabric upon shear arrest, highlight the importance of including fabric tensor as a key parameter in the future development of such constitutive models.

\begin{acknowledgements}
This work was performed, in part, at the Center for Integrated Nanotechnologies, an Office of Science User Facility operated for the U.S. Department of Energy (DOE) Office of Science. Sandia National Laboratories is a multi-mission laboratory managed and operated by National Technology and Engineering Solutions of Sandia, LLC., a wholly owned subsidiary of Honeywell International, Inc., for the U.S. DOE's National Nuclear Security Administration under contract DE-NA-0003525. The views expressed in the article do not necessarily represent the views of the U.S. DOE or the United States Government.
\end{acknowledgements}

Conflict of Interest: The authors declare that they have no conflict of interest.

\section*{Appendix}
In this Appendix, we describe the equations of motion that govern the dynamics of the periodic boundaries that are subjected to an external stress. A modularly-invariant adaptation~\cite{shinoda2004rapid} of the Parrinello-Rahman method~\cite{Parrinello:1981it} of molecular dynamics is utilized to simulate the evolution of a granular system---consisting of $N$ particles with positions and momenta $\{\bm{r}_{i},\bm{p}_{i}\}$ contained within a triclinic periodic box $\bm{H}$ and its associated momentum $\bm{P}_{g}$---under external applied stress $\bm{\sigma}_{a}$. The triclinic periodic box is described by an upper-triangular matrix $H_{ij}=\vec{e}_{i} \cdot \vec{a}_{j}$, where the three lattice vectors $\vec{a}_{j}$ define the periodicity of the triclinic box, and $\vec{e}_{i}$ are the three orthonormal vectors that define the Cartesian coordinate system in the laboratory frame. The equations of motion are given by: 
\begin{eqnarray}
\dot{\bm{r}}_{i}&=&\frac{\bm{p}_{i}}{m_{i}}+\frac{\bm{P}_{g}}{W_{g}}\bm{r}_{i},\\
\dot{\bm{p}}_{i}&=&\bm{F}_{i}-\frac{\bm{P}_{g}}{W_{g}}\bm{p}_{i}-\frac{1}{3N}\frac{\mathrm{Tr\left[\bm{P}_{g}\right]}}{W_{g}}\bm{p}_{i},\\
\dot{\bm{H}}&=&\frac{\bm{P}_{g}}{W_{g}}\bm{H},\\
\dot{\bm{P}_{g}}&=&V\left(\bm{\sigma}-\bm{I}p_{a}\right)-\bm{H}\bm{\Sigma}\bm{H}^{T}+\left(\frac{1}{3N}\sum_{i=1}^{N}\frac{\bm{p}_{i}^{2}}{m_{i}}\right)\bm{I},
\label{eq7}
\end{eqnarray}
where $\bm{F}_{i}$ is the net force on a particle $i$, $V$ is the variable volume of the periodic box, $\bm{I}$ is the identity tensor, and $\bm{\sigma}$ is the internal Cauchy stress that includes contributions from interparticle contact forces and particle momentum (kinetic stress). A `fictitious' mass $W_{g}$ associated with the inertia of the periodic box is set as $W_{g}=Nk_{n}d^{2}/\omega_{g}^{2}$, where $\omega_{g}$ is the frequency of oscillation associated with periodic box fluctuations. The choice of $\omega_{g}$ controls strain rate fluctuations during granular flow. We set $\omega_{g}=0.1\omega_{p}$, where is the $\omega_{p}=\sqrt{m/k_{n}}$ is the frequency associated with the harmonic contact spring between two particles. An additional linear damping is applied to the motion of the periodic box for numerical stability, and its magnitude does not affect the results described here.

The first two terms on the right side of last equation denote the imbalance between internal Cauchy stress and external applied stress that drive the dynamics of the periodic box. The tensor $\bm{\Sigma}$ is defined as:
\begin{equation}
\bm{\Sigma}=\bm{H}_{0}^{-1}\left(\bm{\sigma}_{a}-\bm{I}p_{a}\right)\bm{H}_{0}^{T-1},
\label{eq8}
\end{equation}
where $\bm{H}_{0}$ is the triclinic periodic box matrix  at $t=0$, and $J^{-1}\bm{H}\bm{\Sigma}\bm{H}^{T}$ represents the `true' stress measure of the external deviatoric stress, which is defined with respect to the reference state. Here the Jacobian is $J=\mathrm{det}\left[\bm{F}\right]$, and the deformation gradient is defined as $\bm{F}=\bm{H}\bm{H}_{0}^{-1}$. As a result, this implementation of a constant external stress on the granular system conserves the second Piola-Kirchoff measure of the external stress (\emph{or} equivalently, the thermodynamic tension~\cite{souza1997}). In the present simulations, the reference state is updated to the current state at the end of every time step of integration of the equations of motion, in order to minimize the deviation of internal strain energy from work done by the external stress.

\bibliographystyle{spphys}       % APS-like style for physics
\bibliography{references}   % name your BibTeX data base

%\begin{thebibliography}{}
% etc
%\end{thebibliography}

\end{document}